# Ultrafast changes in lattice symmetry probed by coherent phonons

S. Wall[1*], D. Wegkamp[1], L. Foglia[1], K. Appavoo[2], J. Nag[2], R.F. Haglund Jr.[2], J. Stähler[1], M. Wolf[1]

[1] Fritz-Haber-Institut der Max-Planck-Gesellschaft, Department of Physical Chemistry, Faradayweg 4-6, 14195 Berlin, Germany

[2] Department of Physics and Astronomy, Vanderbilt University, Nashville, Tennessee 37235-1807, USA

**Abstract**

**The electronic and structural properties of a material are strongly determined by its symmetry. Changing the symmetry via a photoinduced phase transition offers new ways to manipulate material properties on ultrafast timescales. However, in order to identify when and how fast these phase transitions occur, methods that can probe the symmetry change in the time domain are required. We show that a time-dependent change in the coherent phonon spectrum can probe a change in symmetry of the lattice potential, thus providing an all-optical probe of structural transitions. We examine the photoinduced structural phase transition in $VO_2$ and show that, above the phase transition threshold, photoexcitation completely changes the lattice potential on an ultrafast timescale. The loss of the equilibrium-phase phonon modes occurs promptly, indicating a non-thermal pathway for the photoinduced phase transition, where a strong perturbation to the lattice potential changes its symmetry before ionic rearrangement has occurred.**

The ultrafast control of a material's properties via non-equilibrium states is a new method to manipulate and control the structural[1], magnetic[2] and electrical[3] properties of materials that may not be accessible under equilibrium conditions. The timescale on which one phase is lost and a new phase emerges is of fundamental interest for understanding how these transitions occur as well as being important for potential applications. However, as these transitions occur via non-equilibrium processes[4], the parameters that define a phase and mark when a transition occurs are difficult to identify.

---

[*] Email: wall@fhi-berlin.mpg.de

In a structural transition, for example, the positions of ions within a crystal are determined by the interactions between the ions and electrons, with the ions residing at minima of the resulting lattice potential energy surface. The lattice potential also determines, through its curvature, the phonon spectrum of the solid, i.e. the response to external perturbations. In most cases, when a solid undergoes a structural phase transition, the symmetry of the lattice potential changes. This modifies both the ionic positions and the phonon spectrum. In equilibrium, these changes are concomitant, however, out of equilibrium the lattice potential may change before the ions have reached their new positions. In such a situation the photoinduced change in the potential defines the onset of the phase transition.

A material's structure can be measured by elastic X-ray or electron diffraction, which maps ionic positions into specific Bragg peaks. Changes in ionic positions are observed by changes to these peaks and, by extending this technique to the time domain, dynamical phase transitions can be clearly indentified[5,6,7,8]. However, the equivalent analysis of the time-resolved changes in the phonon spectrum during a time-dependent phase transition has received less attention. Yet, the evolution of the phonon spectrum may clarify the description of phase transitions, as changes in the potential are responsible for the forces that drive the transition into the new phase.

Phonon spectra can be measured via Raman scattering, which measures the Raman active modes at the $\Gamma$-point of the Brillouin zone. This technique can be extended into the time domain to reveal transient symmetry changes[9], but the time resolution of these experiments is typically low. To overcome this limitation, coherent phonon generation can be used. In an absorbing material, laser light redistributes charge within the solid. This perturbs the lattice potential and produces a force on the ions. The response of ions to the force is determined by the normal modes of the potential, which are expressible as a superposition of phonon modes[10]. If the generated force is fast compared to the period of the phonon, coherent motion of the lattice, or coherent phonons, are induced. This motion can be tracked in the time domain by measuring the effects of the coherent phonons on the transient reflectivity of a material and has been used extensively as a probe to study the lattice in near-equilibrium conditions in the vicinity of phase transitions[11], as well as to look at the phonon softening[12,13,14] and hardening[15] when the system is strongly driven. Phonon softening or stiffening results from a change in the curvature[16,17] of the lattice potential but not from a change in the symmetry of the potential. When the symmetry changes, the number of modes typically changes, which may occur when a soft phonon mode tends to zero frequency in a second order phase transition[14] or may occur step-like in a first order phase transition.

In this article, we demonstrate this principle using the photoinduced phase transition in $VO_2$. A prompt change in the number of modes is observed at high excitation densities, providing an unambiguous marker for the structural phase transition on the ultrafast timescale. These measurements clarify the nature of the photoinduced phase transition mechanism; the pump laser induces, via strong electronic excitations, a prompt change of the symmetry of the lattice potential, which acts as the driving force for the non-thermal transition as evidenced by the ultrafast change in the coherent phonon spectrum. These results demonstrate new ways in which ultrafast structural transitions can be measured through optical probes.

**Results**

**Phonons during the photoinduced phase transition in $VO_2$.** At room temperature, $VO_2$ is a monoclinic $M_1$-phase insulator with $P2_{1/c}$ space group, consisting of four vanadium ions per unit cell. On heating above $T_c$ = 343 K, $VO_2$ becomes metallic and undergoes a first order structural phase transition to a rutile R-phase with the space group $P4_{2/mnm}$ with only two vanadium ions per unit cell. The increased symmetry reduces the number of Raman active phonon modes from 18 in the $M_1$-phase to four in the R-phase without any significant mode softening near the transition temperature[18]. The same structural transition has been induced on the ultrafast timescale by exciting $M_1$-phase $VO_2$ at room temperature with an intense 800 nm pump laser, with a pump fluence greater than $F_{TH}$~ 7 mJ cm$^{-2}$, and observed by femtosecond X-ray[6,19], and electron diffraction[7] as well as through changes in the optical[20,21], and electrical[22,23] properties. These experiments show that the complete transformation to the R-phase, after some initial fast dynamics, is a slow process, taking hundreds of picoseconds to complete. However, measurements of rising edge of the optical transient suggested the phase transition occurred on a timescale limited by the phonon modes of the $M_1$-phase $VO_2$[21].

Figure 1a shows the transient reflectivity of $VO_2$ measured using 800 nm, 40 fs pump and probe for multiple pump fluences at room temperature. Several clear features can be observed. Firstly, there is a large transient that decays back to the initial value with a time constant that depends on fluence. For pump fluences below threshold, oscillations also modulate the transient reflectivity which are the result of the excitation of coherent $M_1$-phase phonons by the pump pulse. Fig. 1b shows a Fourier transform of these oscillations, taken from -300 fs to 4 ps, after subtracting the non-oscillating transient by fitting a two-component exponential decay (see methods section). Four distinct peaks are observed in the spectrum at 4.4, 5.7, 6.7 and 10.2 THz (147, 191, 224, 340 cm$^{-1}$). These frequencies correspond to the four strongest Raman-active phonon modes of the $M_1$-phase that can be coherently excited by our pulse and are in good agreement with those reported in the literature[18].

On increasing the fluence, the amplitudes of these modes increase up to $F_{TH}$, at which point they are reduced and disappear, leaving the decaying transient to dominate the dynamics. This is indicated in Fig. 1c, which shows the Fourier transform of the signal from when excited well above the phase transition threshold, after background subtraction. In this case the signal consists of a single broad feature, and the normal modes of the monoclinic phase cannot be identified.

**Broad band spectroscopy.** The complete loss of the four $M_1$-phase phonons suggests a change in the symmetry of the lattice potential, however, due to the large background it is difficult to determine whether the loss of the phonon modes is instantaneous or due to a dephasing mechanism. To clarify this, we performed broadband probing of the phase transition. Fig. 2a shows 800 nm pump-induced changes in the room temperature reflectivity from 520 nm to 700 nm for below threshold excitation. At all wavelengths coherent oscillations are observed, however, the decaying background is only prominent at longer wavelengths and becomes significantly smaller than the phonon modulation at shorter wavelengths. In line with the observations made with 800 nm probe pulses, the phonons disappear across the whole probed spectrum above the phase transition threshold (Fig. 2b), demonstrating that the loss of phonon-modes is not an optical-probe related artefact. These transient changes in reflectivity are also consistent with the temperature-driven change in reflectivity observed in thin films over the probed region, which show a smaller decrease at shorter wavelengths and a large decrease near 700 nm[24].

Figure 3a shows the transient reflectivity of *p*-polarized 525 nm light after spectral filtering the broadband probe to 5 nm bandwidth, at several pump fluences. At this wavelength, and at low fluences, the transient reflectivity shows a small positive background and is particularly strongly modified by the 5.7 THz phonon, as shown in the Fourier transform of the raw data in Figure 3b. On increasing the pump fluence, the *positive* phonon amplitude increases until the phase transition threshold is reached. At this point the phonon amplitude decreases to zero, and a *negative* transient appears. This is highlighted in Figure 3c, which plots the change in the reflectivity at the peak of the phonon amplitude as well as the reflectivity change of the non-oscillatory signal at 4 ps delay. Also shown in Figure 3c, is the time delay at which a negative reflectivity is first observed (defined as the delay when the reflectivity decrease is larger than 0.1%) which is indicative of the phase transition. This occurs earlier with increasing fluence and, at the highest fluences, occurs during the laser pulse. This, together with the loss of coherent $M_1$-phase phonons indicates that the symmetry of the lattice potential is changed during the excitation process.

Such an ultrafast change in the lattice potential is shown in the schematic of Figure 3d. The initial, low temperature, phase of $VO_2$ is a distorted chain of vanadium ions. This distortion produces multiple optical phonons, simply represented here as $\omega_{L1}$, as well as acoustic phonons, $\omega_{L2}$. Photoexcitation brings the system to the high temperature rutile phase, with a reduced unit cell and a reduced phonon spectrum, represented by a single phonon $\omega_H$, with no Raman active phonons involving vanadium ionic motions. The ultrafast change in the lattice potential is indicated by the middle schematic of Figure 3d, where the laser pulse has changed the lattice potential, and thus the phonon spectrum, before the structure has had significant time to respond. As a result, a displacive excitation is no longer capable of producing coherent phonons.

**Probing the excited state**. At this point it should be noted that the transient dynamics shown in Fig. 1 and Fig. 2 result from multiple processes in addition to coherent phonons, such as the electronic response and coherent interactions between laser beams. These may give rise to additional features (e.g. the kink observed in 20 mJcm$^{-2}$ trace in Fig. 3a), which may not be easily distinguished from coherent phonons. Therefore, to verify that the symmetry change occurs during the excitation process, we perform a pump-probe experiment on the *excited* state of the system below and above the phase transition threshold. Such a technique has been previously used to demonstrate the effects of damping of soft modes[25,26,27].

Figure 4 shows the results at a probe wavelength of 525 nm. Here, the first pulse, $P_1$, excites the system, after some time $\tau_{12}$, a second pump, $P_2$, excites the transient state and the probe pulse measures the combined response, $P_x(t) = P_1(t) + P_2(t-2\pi x/\omega)$, where $\omega$ is the frequency of the phonon mode. Figure 4 shows the transients obtained when the fluence of $P_1$ is below (a) and above (b) the transition threshold and the pump-pump delay is set to be half a period, $P_\pi$ (88 fs), and a full period, $P_{2\pi}$ (176 fs), of the 5.7 THz phonon. The effect of $P_2$ on the time-dependent reflectivity of the excited state can be extracted by subtracting $P_1$ from the measured combined signal, i.e. $P_x - P_1$, which is shown in Figure 4 (c) and (d) for below and above threshold respectively.

In the low fluence regime ($P_1$ = 4.4 mJ cm$^{-2}$, $P_2$ = 1.7 mJ cm$^{-2}$, $P_1 + P_2 < F_{TH}$), the phonon amplitude is switched on and off by the presence of $P_2$, as $P_2$ creates phonons that are either in- ($P_{2\pi}$) or out- ($P_\pi$) of-phase with those generated by $P_1$. This can be seen by the values of the transient reflectivity changes at the dashed lines in Figure 4(c), which are separated by half a phonon period, and show a $\pi$-phase shift. In this case the combined fluence is below the phase transition threshold and the phonon mode does not exhibit a significant red-shift in frequency suggesting that this mode does not significantly soften before the phase transition, as observed in static Raman scattering[18].

The control over the amplitude of the phonon mode is only possible because the symmetry of the lattice potential, and thus the restoring forces, does not change after excitation with $P_1$. However, if $P_1$ is above the threshold and changes the lattice potential on an ultrafast timescale, it should not be possible for $P_2$ to excite coherent phonons that are characteristic of the $M_1$ phase. This is observed in Figure 4c, where $P_1$ is above the phase transition threshold (10 mJ cm$^{-2}$) and $P_2$ is below (3 mJ cm$^{-2}$). As shown in Figure 4d, no correlated response, in- or out-of-phase with $P_1$, is observed in the excited transients, demonstrating that the phonon modes have been lost and no coherent motion can be induced in the excited state of the system. Instead, the second pulse generates an additional large negative transient. The same result was observed when probing at 800 nm, but with a much larger background transient. This shows that the observed peak at approximately 180 fs in the 20 mJcm$^{-2}$ transient of Figure 3c is not the result of a damped oscillation of the $M_1$ phase. Such a feature may correspond to plasmons[28] or even squeezed phonons[29] that cannot be induced by the weaker second pulse. However, at these high fluences, the intensity in the wings of $P_1$, which would otherwise be negligible, may become significant and could also affect the optical signal.

**Discussion**

The loss of all the $M_1$ modes after the 40 fs pump pulse indicates that the entire potential symmetry has changed on an ultrafast timescale. This occurs when a sufficient number of electrons are excited so that their perturbation to the lattice potential becomes large enough to modify the symmetry, subsequently driving the system, non-thermally, into the R-phase.

The interpretation of the ultrafast structural phase transition in $VO_2$ that results from these measurements differs to the previously proposed mechanism in which the timescale was said to be dictated by the phonon period of the monoclinic phase6. On the contrary, at high fluences we observe a complete loss of the coherent motion associated with the monoclinic phase and a drastic change in the response of the excited state, demonstrating a change in symmetry of the lattice potential that is promptly driven by the laser pulse. This analysis brings the ultrafast phase transition mechanism of $VO_2$ into agreement with other Peierls-distorted materials such as bismuth, where the system melts on a sub-phonon-period8. However, unlike bismuth, the transition in $VO_2$ involves complex changes in multiple phonons and results in a final state that is also an ordered solid. It should be emphasized that the establishment of the R-phase with long range order is a slow process, occurring over tens of picoseconds7, that takes much longer than the observed loss in phonon modes6,7,19. The state of $VO_2$ immediately after photoexcitation is far from equilibrium and the lattice potential is likely to continue to evolve to that of the R-phase during the thermalization process[30,31]. On these timescales the evolution of the lattice and the electronic structure may not be concurrent but may be accessible optically through analysis of the broadband dielectric function[32,33].

The analysis presented here is general and can be applied to any ultrafast structural phase transition that changes the symmetry of the Raman tensor, providing an unambiguous ultrafast optical marker for all-optical probing of changes in the lattice potential and its symmetry. In addition, measuring the formation time of the *new* high-symmetry-phase phonon modes that result from the phase transition will provide new information on the establishment of long range order in the system. Thus the study of coherent phonons across photoinduced phase transitions is an ideal complement to time-resolved X-ray measurements. As coherent phonons result from prompt changes in the lattice potential which generate the forces experienced by the ions, they can provide new insight into the mechanisms that *drive* non-equilibrium phase transitions compared to probes of ionic positions.

## Methods

### Samples

Experiments were performed on a 200 nm thin film of $VO_2$ grown on n-doped silicon via pulsed laser deposition[34]. The thermal phase transition was observed at 343 K on heating and the thermal hysteresis was 10 K wide.

### Transient reflectivity

The transient reflectivity was measured at a 50° angle of incidence and performed at room temperature. 800 nm pump and 800 nm probe measurements were performed at laser repetition rate of 150 kHz and measured with a lock-in amplifier. 800 nm pump white-light probe experiments were performed at 100 kHz repetition rate. The duration of the 800 nm pulse was less than 40 fs. The white light pulses were compressed and characterized as described in Ref. 35 to approximately 11 fs duration. The spectrally resolved transient reflectivity was measured directly on a spectrometer allowing the entire spectrum to be obtained at each time delay. Measurements at 525 nm were performed by spectrally filtering the broadband white light pulse to 5 nm bandwidth after reflecting from the $VO_2$ sample and measured on a diode with a lock-in amplifier.

### Fourier transforms of 800 nm data

The Fourier transforms shown in Figures 1b and 1c were performed after subtracting the non-oscillating transient by fitting a background function to the data show in Figure 1a of the form $\Delta R(t)/R = (\text{erf}(t/\tau_p) + 1) *(A_1\exp(-t/\tau_1) + A_2\exp(-t/\tau_2))$, where $\text{erf}(t/\tau_p)$ is the error function and represents the rising edge of the signal, which is dictated by the laser pulse duration $\tau_p$, which is independent of the pump power, $A_i$ and $\tau_i$ ($i$ =1,2) are the fit parameters that represent the amplitude and recovery rate of the dynamics respectively.

**Acknowledgements**

We thank A. Cavalleri for detailed discussions. S.W. acknowledges support from the Alexander von Humboldt Foundation. Research at Vanderbilt was supported by the National Science Foundation (ECCS-0801985).

**Author Contributions**

S.W. conducted the experiments, analyzed the data and wrote the paper. D.W. and L.F. performed measurements. K.A., J.N., R.F.H. grew and characterized the sample. S.W., D.W., L.F., R.F.H., J.S. and M.W. discussed the results and interpretation, and the manuscript.

The authors declare no competing financial interests.

**Figure captions**

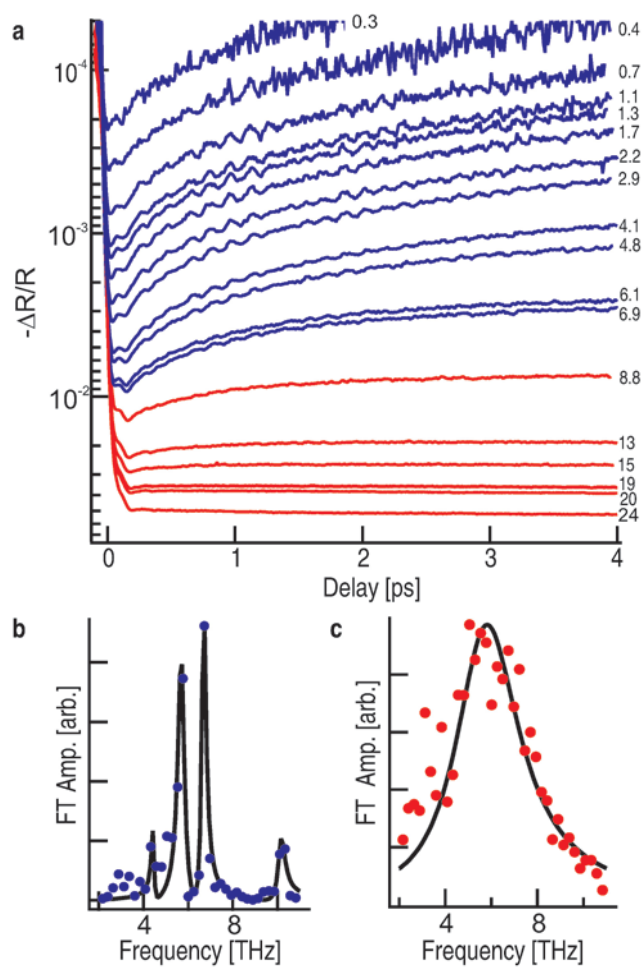

**Figure 1 | Dynamics of the reflectivity at 800 nm across the photoinduced phase transition.**

(**a**) Transient reflectivity at 800 nm as a function of pump fluence. The numbers to the right of each transient indicate the pump fluence measured in mJcm$^{-2}$. Blue lines correspond to transients below the 7 mJ cm$^{-2}$ threshold, red lines above. Fourier transforms (FT) of the transient reflectivity after subtracting a fitted background (see method section for details) for below threshold (1.7 mJcm$^{-2}$) (**b**) and above (24 mJcm$^{-2}$)(**c**). Below threshold the four lowest Raman active modes are clearly observed (solid line indicates the best fit to four Lorentzian oscillators). Above threshold all $M_1$ phonons are lost and a single broad feature is observed.

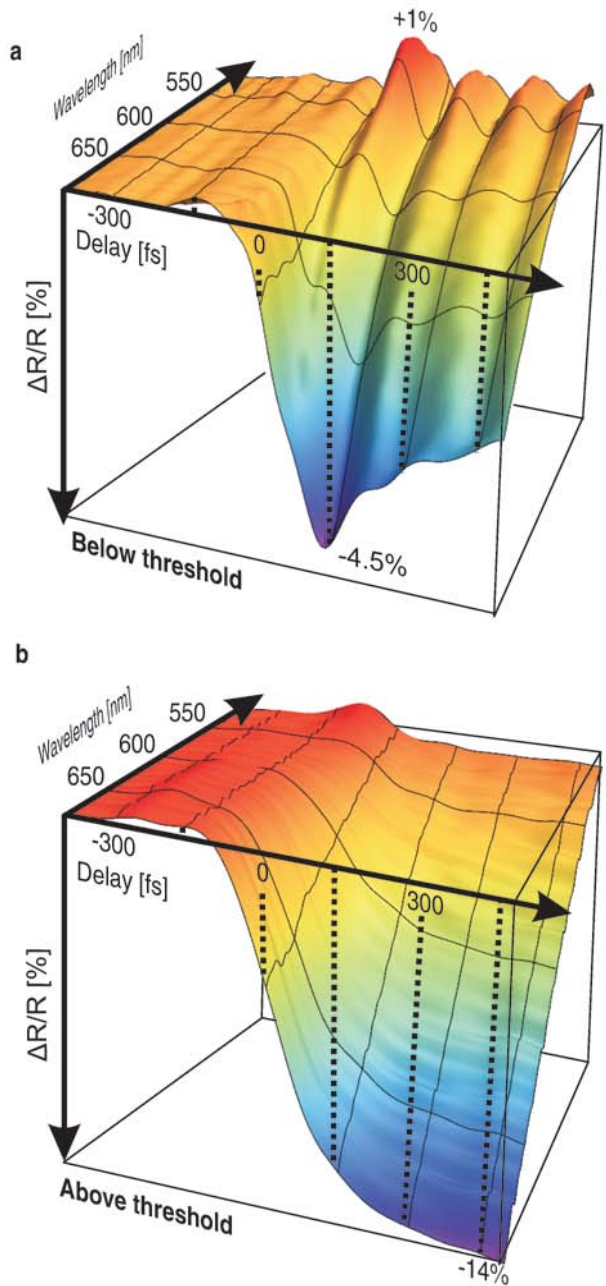

**Figure 2 | Broadband transient spectroscopy of the photoinduced phase transition.**

Transient reflectivity as a function of probe wavelength and delay, (**a**) below and (**b**) above the photoinduced phase transition threshold. Below threshold, signals at longer wavelengths are dominated by a large negative transient with the oscillations dominating the signal at shorter wavelengths. Above threshold, no oscillations are observed, showing that the phonon modes are lost over the entire visible spectrum.

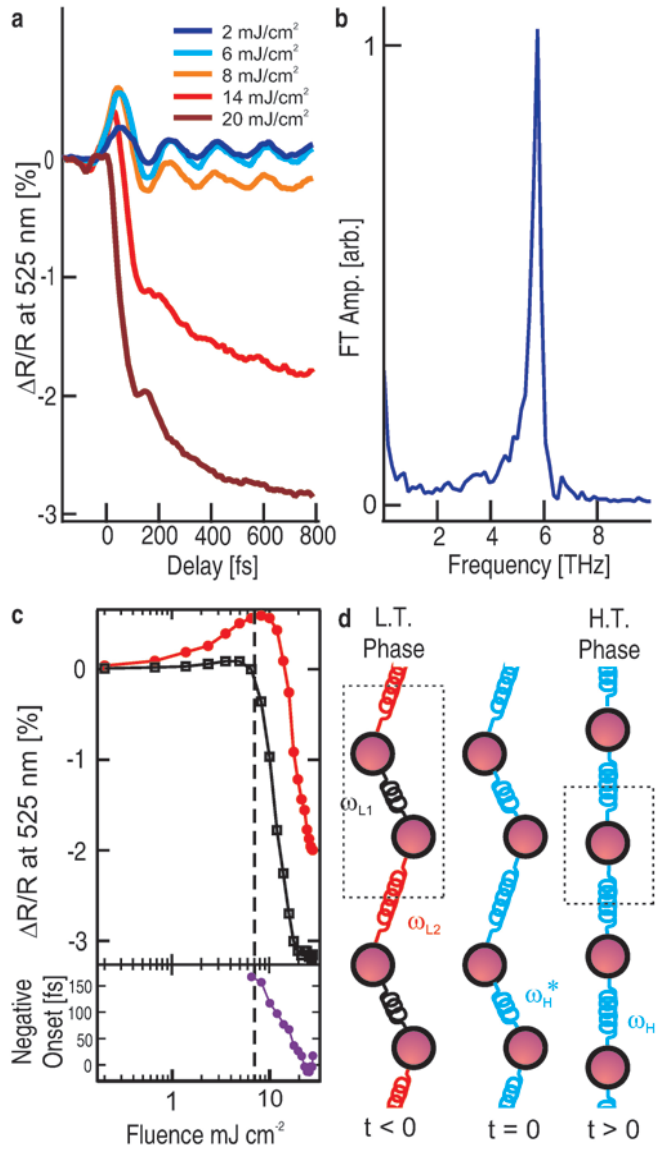

**Figure 3 | Evolution of the phonon modes during the photoinduced phase transition.**

(**a**) The evolution of the reflectivity at a probe wavelength of 525 nm for various fluences. (**b**) The Fourier transform of the 2 mJ cm$^{-2}$ transient without background subtraction. (**c**) Fluence dependence of peak signal measured at 60 fs (red circles) and the background level, measured at 4 ps (black open squares) demonstrating that the phonon amplitude increases until the transition threshold (dashed line at 7 mJ cm$^{-2}$) at which point the negative transient dominates. Also plotted (purple circles) is the time after excitation at which a negative change in reflectivity is observed (defined as when the reflectivity change drops below < -0.1%). At the highest fluences the onset of the negative transient is limited by the pulse duration. (**d**) Simplified schematic of the phase transition in VO$_2$. Low temperature (L.T.) and high temperature (H.T.) vanadium positions are shown, which are also the positions before (*t*<0) and after (*t*>0) excitation. The low and high temperature phases can be characterized by a different set of spring constants $\omega_x$, which are promptly changed after excitation. At $t$=0 the lattice potential is changed before significant atomic motion. Dashed boxes represent the unit cells.

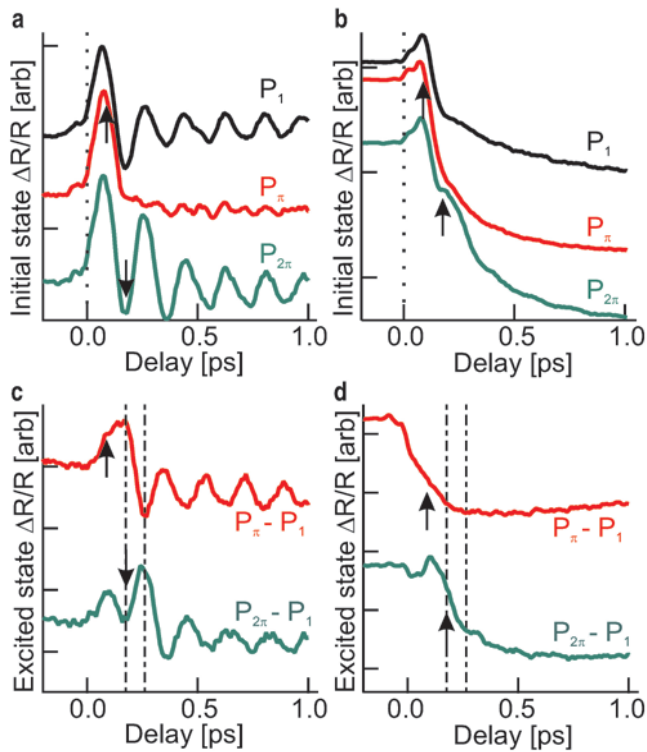

**Figure 4 | Probing the response of the excited state.**

Pump-pump-probe measurements (**a**) below threshold and (**b**) above. Traces labelled $P_1$ correspond to the transient reflectivity generated by a single pump pulse. Traces labelled $P_\pi/P_{2\pi}$ correspond to a double pulse excitation when the pump-pump delay is set to be out of phase, $P_\pi$, or in phase, $P_{2\pi}$, with the 5.7 THz phonon mode. The arrows indicate the arrival time of the second pump pulse. (**c**) and (**d**) correspond to the transient response of the excited state below and above threshold respectively obtained by subtracting the single pulse transient reflectivity from the double pulse transient reflectivity. Dashed lines correspond to the points separated by half of the 5.7 THz phonon period, which are anti-correlated below threshold and exhibit no correlation above threshold.